\documentclass[twocolumn,twoside,slac_two,floatfix]{revtex4}
\usepackage{graphicx}
\usepackage{fancyhdr}
\begin{document}

\title{Connection between Gamma-Ray Variations and Disturbances in the Jets of Blazars}

%

\author{S. Jorstad, A. Marscher, F. D'Arcangelo, and B. Harrison}
\affiliation{Boston University, 725 Commonwealth Ave., Boston, MA 02215, USA}

\begin{abstract}
We perform monthly total and polarized intensity imaging of a sample of $\gamma$-ray blazars (33 sources) with the Very Long Baseline Array (VLBA) at 43 GHz with the high resolution of 0.1 milliarcseconds. From Summer 2008 to October 2009 several of these blazars triggered Astronomical Telegrams due to a high $\gamma$-ray state detected by
the Fermi Large Area Telescope (LAT): AO 0235+164, 3C 273, 3C 279, PKS 1510-089, and 3C 454.3. We have found that 1) $\gamma$-ray flares in these blazars occur during an increase of the flux in the 43 GHz VLBI core; 2) strong $\gamma$-ray activity, consisting of several flares of various amplitudes and durations (weeks to months), is simultaneous with
the propagation of a superluminal knot in the inner jet, as found previously for BL Lac (Marscher et al. 2008); 3) coincidence of a superluminal knot with the 43 GHz core precedes the most intense $\gamma$-ray flare by 36$\pm$24 days. Our results strongly support the idea that the most dramatic $\gamma$-ray outbursts of blazars originate in the vicinity of the mm-wave core of the relativistic jet. These results are preliminary and should be tested by future monitoring with the VLBA and Fermi.
\end{abstract}

\maketitle

\thispagestyle{fancy}

\section{Introduction} 
During the EGRET era, \citet{J01a,J01b} undertook an extensive study   
of changes in the structure of the compact radio jets of EGRET-detected blazars with the Very Long Baseline Array (VLBA) at 43 and 22~GHz in order to investigate correlations between the properties of the compact radio jets and $\gamma$-ray emission. They found that the $\gamma$-ray emission originates in a highly relativistic jet and that the distribution of apparent velocities of jet components peaks at 10-12~c, which is significantly higher than the average superluminal speed of jet components in the general population of strong compact radio sources. They also found a positive correlation between VLBI core flux and $\gamma$-ray flux. These findings imply that the $\gamma$-rays are more highly beamed than the radio synchrotron radiation. Recently the results received some confirmation with a much larger sample of compact radio sources and dramatically improved $\gamma$-ray data provided by the Fermi Large Area Telescope (LAT) \citep{LIST09,KOV09}. However, the location(s) of the origin of $\gamma$-ray emission as well as the mechanism(s) of its production remain unclear. The conclusion that the $\gamma$-ray emission depends more strongly on the Lorentz factor of the jet flow than does the radio emission is consistent with both synchrotron self-Compton and external inverse Compton models, as 
well as with non-inverse Compton models in which the $\gamma$-ray emission arises in the high-radiation environment close to the central engine. Although
\citet{J01b} found a  statistically significant connection between ejections of superluminal components from the VLBI core and high states of $\gamma$-ray emission, the EGRET $\gamma$-ray light curves were sparse and the VLBA monitoring had significant gaps that caused identifications of knots across epochs to be non-unique.  
    
We are currently executing a comprehensive program of multifrequency monitoring of a sample of 33 blazars detected in $\gamma$-rays either by EGRET \citep{HART99} or by the LAT from August to October 2008 \citep{A09a}
(except 2 quasars, 3C~345 and 3C~446, which were not yet known to be $\gamma$-ray bright). Each source is observed 1) with the VLBA at 43~GHz monthly (some bimonthly),
2) at optical wavelengths, including polarization with different telescopes
as often as possible, and 3) at X-rays with RXTE and Swift. In this proceeding we discuss some preliminary results obtained from analysis of the VLBA images along with the LAT $\gamma$-ray light curves.

\section{Data Reduction}
We processed the VLBA data and created images in a manner identical to
that described in \citet{J05}. The VLBA images of all 33 blazars are available at our website http://www.bu.edu/blazars/VLBAproject.html. We modelled the images in terms of a small
number of components with circular Gaussian brightness distributions. The core is a stationary feature located at one of the end of the portion of the jet that is visible at 43~GHz. Identification of components in the jet across the epochs is based on analysis of
their flux, position angle, distance from the
core, polarization, and size. Figures 1-4
show sequences of the total and polarized intensity images of 0235+164, 1510-089, 3C~273, 3C~279, and 3C~345. The knot(s) of total and/or polarized flux enhancement moving down the jet are pronounced in these sequences. The speed and time of ejection of each knot were calculated from a ballistic part of its trajectory, with H$_\circ$=71 km s$^{-1}$Mpc$^{-1}$ and the
concordance cosmology (Spergel et al. 2007).  Figure 5 shows the light curves of the core (red) and separation vs.
time of moving knots with respect to the
core.  We use $\gamma$-ray flux measurements posted on the LAT website for  0235+164, 3C~273, 3C~279, 
1510-089, 3C~454.3 and 0528+164 with a bin size of one week. We have calculated the $\gamma$-ray light curve for 3C 345 with the same binning interval 
using the LAT data and the threads provided
on the Fermi website. We modeled the $\gamma$-ray emission of the quasar by a single power-law model with a constant photon index $-$2.1. A single power-law model was used also
to determine the contribution of bright $\gamma$-ray sources within 15$^\circ$ from 3C~345:  1641+3939, Mrk501, and 1633+382, for which both prefactor and photon index were calculated. The $\gamma$-ray light curves are presented in Figure 5 by a blue  color. 

\section{Discussion and Conclusions}
We have performed a Kendall $\tau$ cross correlation analysis \citep{AstroStat}  between the VLBI core and $\gamma$-ray light curves.
This method is applicable even if the data contain upper or lower limits for some measurements. We find a positive correlation in all 6 cases, which is significant at a confidence level of 99\% in the case of 0235+164, 
1510-089, 3C~454.3, and 0528+134. Brightening of the VLBI core is usually caused by a disturbance propagating through the core \citep{SAV02}. The disturbance is seen in VLBA images as a knot of enhanced total and polarized intensity (Figs. 1-4). Table 1 gives the coefficients of correlation between the VLBI core and $\gamma$-ray light curves plus the relative timing between $\gamma$-ray peaks and ejections of superluminal
components from the core. The columns are as follows: 1 - source name; 2 - Kendall coefficient of correlation between the VLBI core and $\gamma$-ray light curves, $\tau$; 3 - knot designation; 4 - time of the ejection of the superluminal knot from the core at 43 GHz, $T_\circ$ in RJD, where RJD=JD-2450000; 5 - time of the peak of the $\gamma$-ray flare corresponding to the epoch of a local maximum in the $\gamma$-ray light curve, $T_\gamma$; 6 - $\Delta T$ = $T_\circ$ - $T_\gamma$; 7 - apparent speed of the superluminal knot in units of c, $\beta_{app}$; and 
8 - the flux peak of the $\gamma$-ray flare, $S_\gamma^{max}$.  
 
\begin{table*}[t]
\begin{center}
\caption{Correspondence between Gamma-Ray Flares and Time of Ejections of Superluminal Knots}
\begin{tabular}{|l|l|l|r|r|r|r|r|}
\hline \textbf{Source} & \textbf{$\tau$} & \textbf{Knot} &
\textbf{$T_\circ$} & \textbf{$T_\gamma$} & \textbf{$\Delta T$} & \textbf{$\beta_{app}$} & \textbf{$S_\gamma^{max}$} \\ 
\text{} & \text{} & \text{} &
\text{RJD} & \text{RJD} &\text{days} & \text{c} & \text{10$^{-6}$ph cm$^{-2}$s$^{-1}$} \\ 
\hline 0235+164& 0.79 & K1 & 4728$\pm$30 & 4730$\pm$4 & -2$\pm$34 &56$\pm$10&0.91$\pm$0.07  \\
\hline 3C 273& 0.21 & K2 & 4747$\pm$45 & 4744$\pm$4 & +3$\pm$49 &8.3$\pm$1.4&1.40$\pm$0.13  \\
&  & K3 & 4901$\pm$33 & 4947$\pm$4 & -46$\pm$37 &8.9$\pm$0.7&1.03$\pm$0.10  \\
&  & K4 & $\sim$5029 & 5100$\pm$4 & $\sim$-71 &$\sim$23&5.13$\pm$0.27  \\
\hline 3C 279 & 0.32 & K2 & 4779$\pm$24 & 4800$\pm$4 & -21$\pm$29 &14.5$\pm$2.0&1.48$\pm$0.10  \\
\hline 1510$-$089& 0.56 & K1 & 4675$\pm$23 & 4723$\pm$4 & -48$\pm$27 &24.0$\pm$2&0.91$\pm$0.07  \\
&  & K2 & 4959$\pm$4 & 4962$\pm$4 & -3$\pm$8 &21.6$\pm$0.6&3.78$\pm$0.17  \\
\hline 3C 345& 0.28 & K1 & 4677$\pm$21 & 4737$\pm$4 & -60$\pm$25 &7.1$\pm$0.6&0.15$\pm$0.07  \\
&  & K2 & 4904$\pm$50 & 4982$\pm$4 & -78$\pm$54 &10.2$\pm$2.2&0.28$\pm$0.09  \\

\hline
\end{tabular}
\label{tab1}
\end{center}
\end{table*}
The case of the quasar 3C~454.3 is more complex (Jorstad et al. 2010, submitted to ApJ). Despite the strong correlation between the VLBI core and $\gamma$-ray variations ($\tau$=0.69), the variability observed in 2008 (RJD: 4700-4800) is significantly delayed with respect to the ejection of a superluminal component (by $\sim$200 days) that can be associated with the increase of flux in the VLBI core. However, that ejection did coincide with a very strong and fast optical outburst. Such a rapid, high-amplitude flare is consistent with the high Doppler factor derived for the knot, $\delta\sim$25. The prolonged period of elevated radio flux in the core, along with high $\gamma$-ray flux, suggest that the core is extended, with multiple sites of particle acceleration. For the second $\gamma$-ray outburst, which started in 2009 Autumn, it is too early to discuss what is going on in the core, which is brightening dramatically.

Although the number of events is small, this sample of blazars includes all sources from our monitoring list that exhibited strong $\gamma$-ray flares during the first year of Fermi LAT operation. Analysis of results given in Table 1 suggests that the ejection of a superluminal knot from the mm-wave core precedes a high $\gamma$-ray state by $\sim$ a month, in agreement with previous findings by \citet{J01b, LV03}. Figure 5 shows that, for a number of $\gamma$-ray flares, the fading branch of a $\gamma$-ray outburst coincides with that of the VLBI core. This implies that a high $\gamma$-ray state ends when a disturbance completely leaves the mm-wave core region. 

The $\gamma$-ray variations of 3C~454.3 correlate very well with the optical behavior \citep{BON09}. This suggests that $\gamma$-ray emission is generated by the inverse-Compton mechanism, with the relativistic electrons that produce optical synchrotron emission in the jet scattering seed photons from either inside the jet (SSC) or external to the jet (EC). In the case of the EC process, the seed photons should be local to the mm-wave VLBI core.

Our results show that the mm-VLBI core is a very important feature in blazar jets that serves as a unique reference point for determining the location in the jet of events that we see in high-energy light curves.  VLBI images allow detailed descriptions of the events in this portion of the jet if intense high-resolution monitoring  is performed. Currently, the VLBA is the only instrument that can provide such monitoring and, therefore, advance our understanding of the physics of the relativistic jets in blazars.
       
\begin{acknowledgments}
This research was funded in part by NASA through Fermi Guest
Investigator grants NNX08AV65G and NNX08AV61G, and by the National
Science Foundation through grant AST-0907893.
\end{acknowledgments}

\clearpage
\begin{figure*}
\includegraphics[width=155mm]{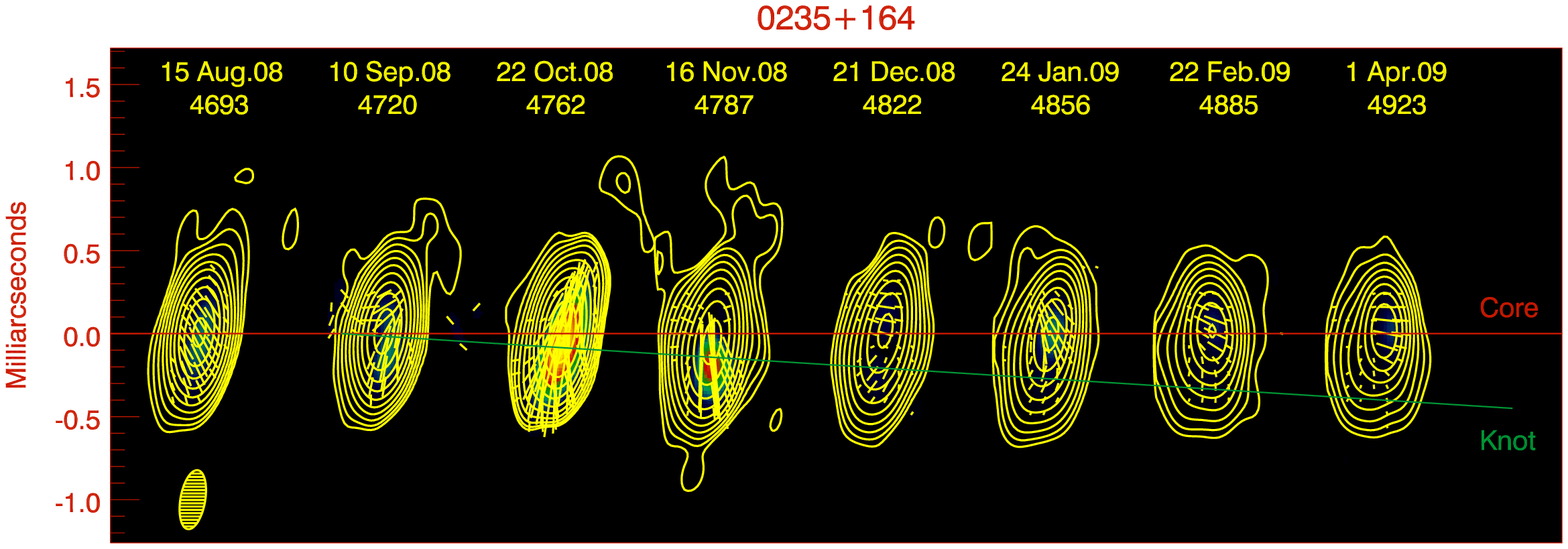}
\caption{Total (contours) and polarized (color scale) intensity images of    0235+164; yellow line segments show direction of polarization; the size of the beam is 0.36$\times$0.15 mas$^2$ at PA=-10$^\circ$, S$_{peak}$= 4.0 Jy/beam, S$_{pol}$= 143 mJy/beam, contours are 0.25, 0.5, É64\% of the total intensity peak.}
\label{0235vlba}
\end{figure*}

\begin{figure*}
\includegraphics[width=155mm]{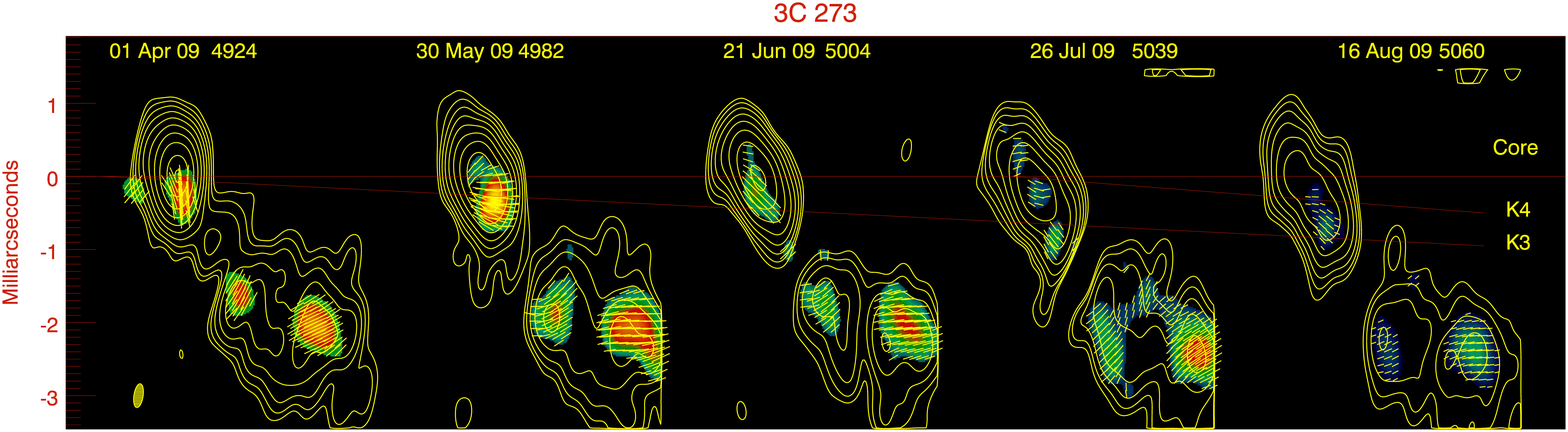}
\caption{Total (contours) and polarized (color scale) intensity images of    3C~273; yellow line segments show direction of polarization; the size of the beam is 0.38$\times$0.14 mas$^2$ at PA=-10$^\circ$, S$_{peak}$= 8.31 Jy/beam, S$_{pol}$= 86 mJy/beam, contours are 0.25, 0.5, É64\% of the total intensity peak.}
\label{3c273vlba}
\end{figure*}

\begin{figure*}
\includegraphics[width=155mm]{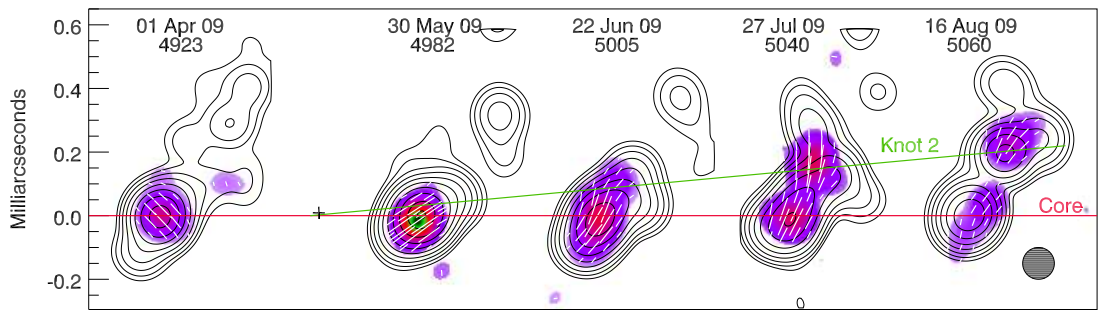}
\caption{Total (contours) and polarized (color scale) intensity images of 1510-089; white line segments show direction of polarization; the size of the beam is 0.1$\times$0.1 mas$^2$, S$_{peak}$= 3.1 Jy/beam, S$_{pol}$= 120 mJy/beam, contours are 0.5, 1, É64\% of the total intensity peak.}
\label{1510vlba}
\end{figure*}

\begin{figure*}
\includegraphics[width=155mm]{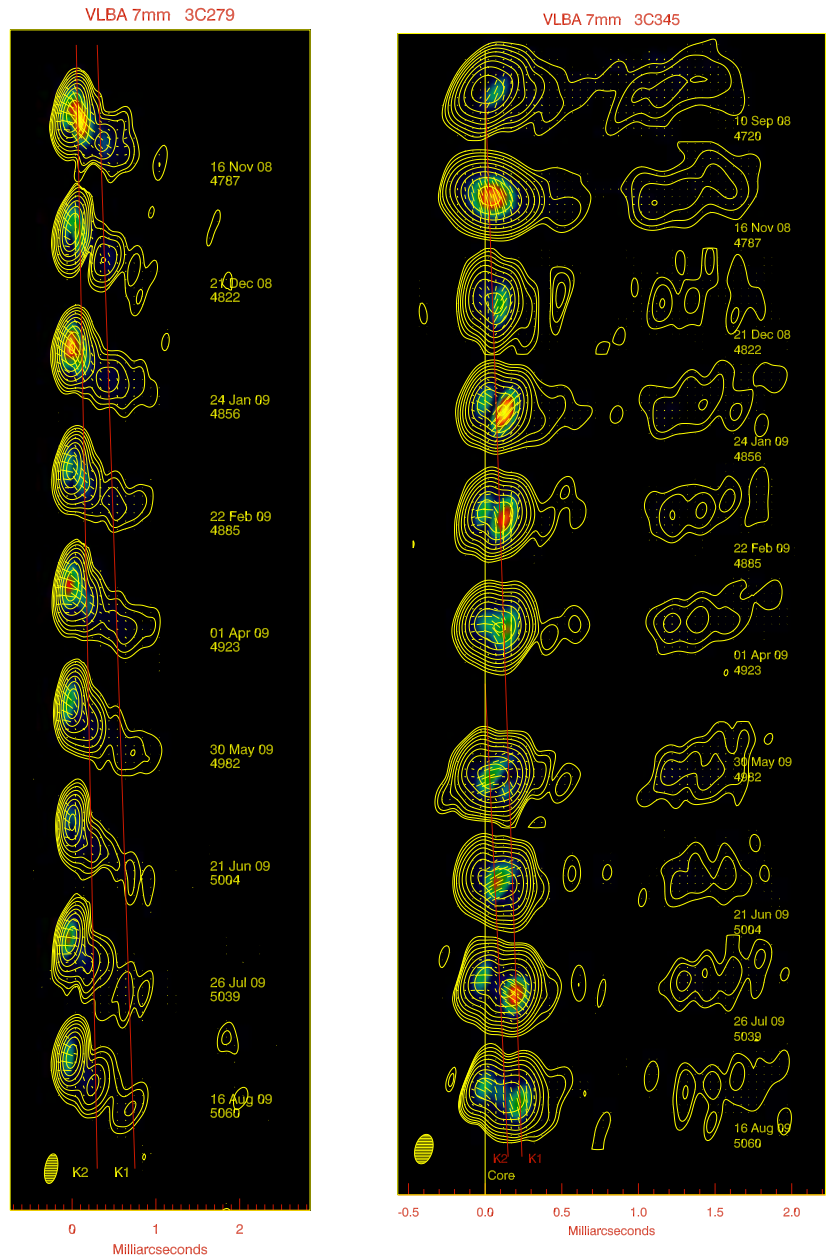}
\caption{Total (contours) and polarized (color scale) intensity images of the quasars 3C~279 ({\it left}) and 3C~345 ({\it right}); yellow line segments show direction of the polarization; 3C 279: beam= 0.36$\times$0.15 mas$^2$  at PA=-10$^\circ$, S$_{peak}$=12.0 Jy/beam, S$_{pol}$=200 mJy/beam; 3C 345: beam= 0.25$\times$0.14 mas$^2$  at PA=-10$^\circ$, S$_{peak}$=3.3 Jy/beam, S$_{pol}$=210 mJy/beam; contours are 0.25, 0.5, É64\% of the total intensity peak.}
\label{3c279vlba}
\end{figure*}

\begin{figure*}
\includegraphics[width=155mm]{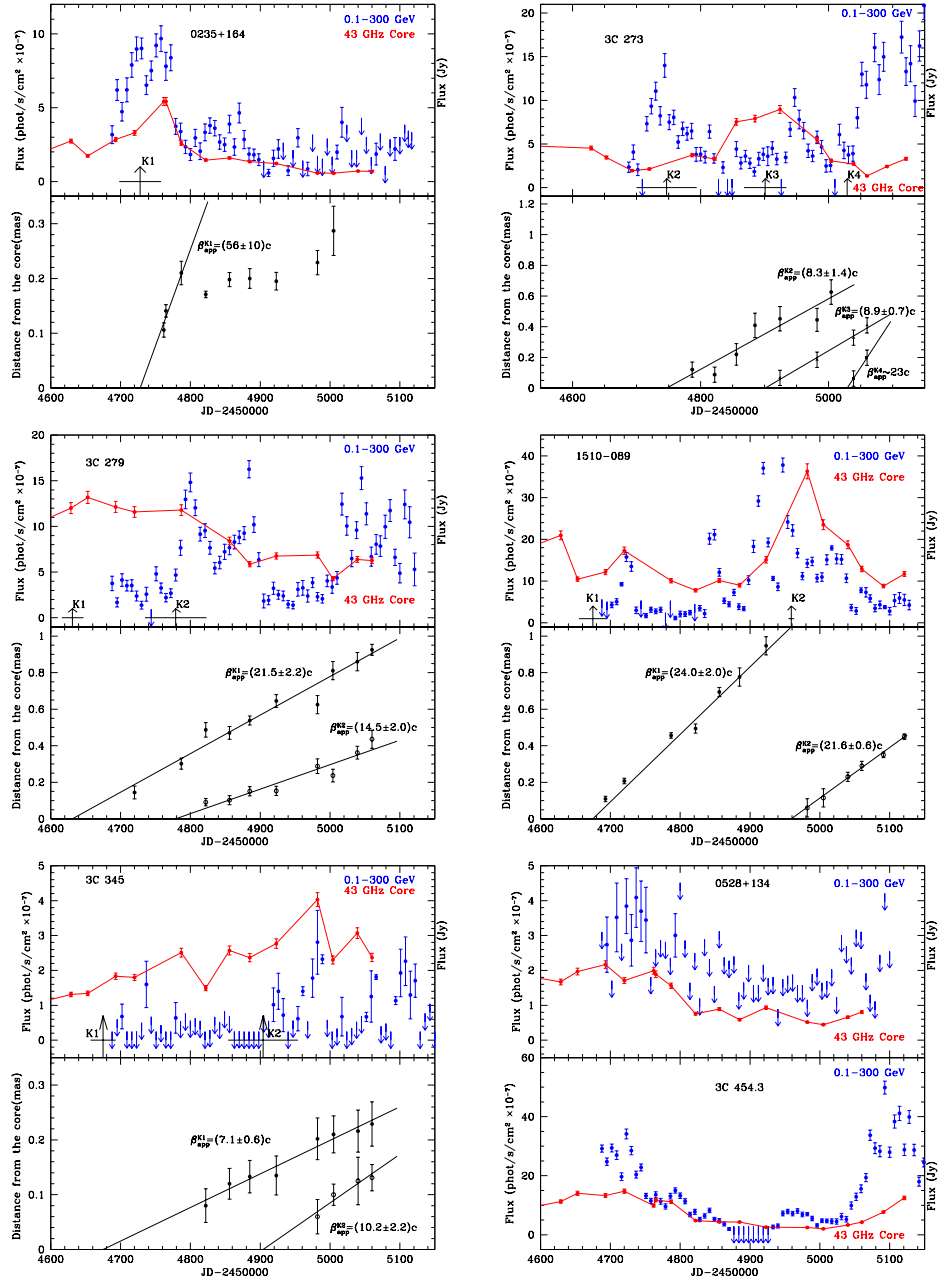}
\caption{Gamma-ray (blue) and 43 GHz VLBI core (red) light curves of 0235+164, 3C~273, 3C~279, 1510-089, 3C~345, 0528+134, and 3C~454.3 ; arrows indicate the time of ejection of superluminal knots  from the core; the time of the ejections is determined from a linear fit to the motion of knots near the core, as shown at the bottom  plot of panels for 0235+164, 3C~273, 3C~279, 1510-089, and 3C~345.}
\end{figure*}
\end{document}